\documentclass[12pt]{amsart}

\usepackage[utf8]{inputenc}
\usepackage[T1]{fontenc}
\usepackage{amssymb,amsfonts}
\usepackage{mathtools}
\usepackage{booktabs}
\usepackage{microtype}
\usepackage{url}
\usepackage{hyperref}
\usepackage{graphicx}
\usepackage{algorithm}
\usepackage{algpseudocode}

\allowdisplaybreaks
\makeatletter
\@namedef{subjclassname@2020}{%
  \textup{2020} Mathematics Subject Classification}
\makeatother

\hypersetup{
  colorlinks=true,
  linkcolor=blue,
  citecolor=blue,
  urlcolor=blue
}

\newtheorem{theorem}{Theorem}[section]
\newtheorem{lemma}[theorem]{Lemma}
\newtheorem{proposition}[theorem]{Proposition}
\newtheorem{corollary}[theorem]{Corollary}

\theoremstyle{definition}
\newtheorem{definition}[theorem]{Definition}
\newtheorem{example}[theorem]{Example}

\theoremstyle{remark}
\newtheorem{remark}[theorem]{Remark}

\numberwithin{equation}{section}

\newcommand{\Q}{\mathbb{Q}}
\newcommand{\R}{\mathbb{R}}
\newcommand{\N}{\mathbb{N}}
\newcommand{\Nzero}{\mathbb{N}_0}
\newcommand{\Z}{\mathbb{Z}}
\newcommand{\Zp}{\mathbb{Z}_p}
\newcommand{\Qp}{\mathbb{Q}_p}
\newcommand{\abs}[1]{\lvert #1\rvert_p}
\newcommand{\rabs}[1]{\lvert #1\rvert}
\newcommand{\expp}{\exp_p}
\newcommand{\TV}{\operatorname{TV}}
\newcommand{\tA}{\widetilde A}
\newcommand{\tf}{\widetilde f}

\DeclareMathOperator*{\argmin}{arg\,min}
\DeclareMathOperator*{\argmax}{arg\,max}

\title[Certified readout for p-adic aggregation]
{Certified finite-precision readout for \texorpdfstring{$p$-adic}{p-adic} kernel aggregation}

\author{Su Hu}
\address{Department of Mathematics, South China University of Technology, Guangzhou, Guangdong 510640, China}
\email{mahusu@scut.edu.cn}

\author{Min-Soo Kim}
\address{Department of Mathematics Education, Kyungnam University, Changwon, Gyeongnam 51767, Republic of Korea}
\email{mskim@kyungnam.ac.kr}


\begin{document}

\begin{abstract}
In non-archimedean numerical analysis, a verified $p$-adic approximation can determine an absolute value, and hence downstream probabilistic rankings and classification decisions, with mathematical exactness long before the underlying number is fully resolved. We turn this observation into rigorous finite-precision certificates for norm-based readouts of amplitude families. For a common error radius, a nonempty resolved set already fixes every exact maximizer, ties included, at the sharp integer absolute precision $v_*+1$, and conditioning on that set yields an exact total-variation identity separating unresolved $p$-adic mass from real rounding error. For aggregates of $p$-adic exponentials, we give a centered-moment certificate with an explicit remainder bound: on rational data it runs in exact rational arithmetic without individual exponential evaluations, attains its minimal bound at an active phase, and resolves every nonzero aggregate at high order; we also obtain an aggregate perturbation bound free of summand-count factors. For multiclass $p$-adic classifiers that decode by minimizing the norm of a class discriminant, we prove a necessary and sufficient certification condition and show that the sharp common precision threshold is $v^{(2)}+1$, governed by the second-largest discriminant valuation, so the winning discriminant need not itself be resolved. We apply these certificates to discrete $p$-adic path-integral actions on uniform affine chains, deriving explicit action centers and proving exact output-independence at zero effective quadratic coupling. Pseudocode and a reproducible reference implementation accompany the theory.
\end{abstract}
\maketitle

%=====================================================================
\section{Introduction}
%=====================================================================

Throughout this paper, $p$ denotes an odd prime, $\abs{p}=p^{-1}$,
$\N=\{1,2,\ldots\}$, and $\Nzero=\N\cup\{0\}$.
Throughout, the absolute-value notation $\abs{\cdot}$ denotes the $p$-adic
absolute value $\lvert\cdot\rvert_p$; the archimedean (real) absolute value is
written $\rabs{\cdot}$ and is used only for real quantities such as
total-variation distance.
The output space $Y$ is a finite, nonempty set equipped with a fixed total
ordering.

\subsection*{Motivation: Archimedean vs.\ non-archimedean certification}

Numerical implementations of probabilistic models and kernel classifiers must control two distinct quantities: the approximation error of the reported probability distribution and the correctness of the discrete decision (such as top-$1$ classification or argmax decoding). In standard archimedean (IEEE 754 floating-point) arithmetic, backward error analysis and interval computation provide the foundational toolkit \cite{Hig,BHH,MKC,Rum}. However, when real weights are known only up to positive-radius error intervals, certifying a strict maximizer requires a positive separation gap between the leading intervals, while certifying exact ties is impossible in general without symbolic side information---a phenomenon closely analogous to the Table Maker's Dilemma in verified floating-point rounding. Furthermore, when aggregating $M$ exponential terms $\sum_{j=1}^M \alpha_j \exp(z_j)$ in real arithmetic, worst-case rounding error bounds typically scale with the summand count $M$ and suffer from catastrophic cancellation when phases interfere destructively.

In the non-archimedean ($p$-adic) setting, numerical computations arise naturally in two rapidly developing areas:
\begin{enumerate}
\item \emph{$p$-adic machine learning and hierarchical representation} \cite{Martins}, where data with tree-like or ultrametric structure are embedded into $\Qp$ and classified by comparing $p$-adic norms of linear or kernel discriminants;
\item \emph{$p$-adic mathematical physics and discrete path integrals} \cite{VVZ,DKKV,Koc,HuKim2026}, where transition amplitudes are formed by aggregating $p$-adic exponential kernels over discrete state spaces and physical observables are extracted from powers of their $p$-adic norms $\abs{A_y}^\beta$.
\end{enumerate}
While general $p$-adic precision tracking via ultrametric balls and differential precision lattices is well established \cite{CRV,Car}, standard $p$-adic arithmetic treats the amplitude $A_y\in\Qp$ itself as the target of computation. Yet in both learning and physical modeling, the final observable is a \emph{norm-based readout}.

The starting point of this work is the elementary ultrametric stability property: if a verified approximation $\tA\in\Qp$ to $A\in\Qp$ satisfies
$$
\abs{A-\tA}\le\delta
\quad\hbox{and}\quad
\abs{\tA}>\delta,
$$
then the strong triangle inequality forces
\begin{equation}\label{eq:key}
\abs{A}=\abs{\tA}.
\end{equation}
Consequently, a numerical readout depending only on $p$-adic absolute values can become \emph{mathematically exact} at low finite precision, even while arbitrarily many higher $p$-adic digits of the underlying amplitudes $A_y$ remain unknown.

\begin{remark}[Discrete visibility and boundary-free certification]\label{rem:discrete}
The $p$-adic norm takes values in $p^{\Z}\cup\{0\}$ and is locally constant on $\Qp^\times$, so any norm clearing its verified radius $\delta$ is pinned down with zero error. In real arithmetic a readout factoring through a discrete step function (e.g.\ the binary exponent $\lfloor\log_2\rabs{x}\rfloor$) is discontinuous at every dyadic boundary $2^n$, and no positive-width interval straddling $2^n$ can certify it. Ultrametric balls are open and closed and partition $\Qp^\times$ without boundary points, so the $p$-adic norm has no boundary ambiguity outside $0$; comparisons with real arithmetic below refer to readouts depending continuously on real weights (such as real softmax \cite{BHH}).
\end{remark}

\subsection*{Readout-level certification and the error ledger}

For an amplitude family $A=(A_y)_{y\in Y}\in\Qp^Y$, not identically zero, inverse temperature $\beta>0$, and positive real prior weights $h_y>0$, we study the normalized readout
\begin{equation}\label{eq:readout-intro}
P_{\beta,h}(y\mid A)
=
\frac{h_y\abs{A_y}^{\beta}}
{\sum_{z\in Y}h_z\abs{A_z}^{\beta}}.
\end{equation}
Given verified representatives $(\tA_y,\delta)$ with a common error radius $\delta\ge0$, we partition the output space into the \emph{resolved set} and \emph{unresolved set}:
$$
I=\{y\in Y:\abs{\tA_y}>\delta\},
\qquad U=Y\setminus I.
$$
When $h\equiv1$ and $I\ne\varnothing$, every exact maximizer of $\abs{A_y}$ is guaranteed to lie in $I$, and all norms on $I$ are known exactly. Consequently, the complete maximizer set---including exact ties, which occur frequently due to the discrete value group $p^\Z$---is certified without gap assumptions. Moreover, setting $S_I=\sum_{y\in I}\abs{\tA_y}^{\beta}$, the conditional distribution $Q=P_\beta(\cdot\mid I)$ is computable in exact rational arithmetic (when $\beta\in\N$) and satisfies the exact total-variation identity
$$
\TV(P_\beta,Q)=P_\beta(U)
\le
\frac{\vert{}U\vert{}\delta^\beta}{S_I+\vert{}U\vert{}\delta^\beta}.
$$
When $Q$ is subsequently converted to a floating-point or normalized dyadic vector $\widehat Q$, our framework provides an explicit \emph{error ledger} (Proposition~\ref{prop:real}) that decouples the non-archimedean truncation mass $P_{\beta,h}(U)$ from the archimedean output rounding error $\TV(\widehat Q,Q)$.

\subsection*{Exponential aggregation and centered-moment certificates}

Our principal computational target is the kernel aggregate
\begin{equation}\label{eq:aggregate}
A_y=\sum_{j\in J_y}\alpha_j\expp(z_{yj}),
\qquad z_{yj}\in p\Zp,
\end{equation}
where $J_y$ is the active source set at output $y$ (with $J_y=\varnothing$ yielding $A_y=0$ exactly). Because $\expp$ is an isometry on $p\Zp$, we first show that direct modular evaluation enjoys a perturbation bound governed by the \emph{maximum} source error rather than a sum over $\vert{}J_y\vert{}$ sources (Theorem~\ref{thm:aggregate}).

When severe algebraic cancellation occurs among the coefficients $\alpha_j$ (for instance, in balanced zero-sum configurations $\sum_j\alpha_j=0$), direct evaluation requires expanding every individual exponential $\expp(z_{yj})$ to high modular precision $p^N$. To bypass individual exponential evaluations, we introduce a \emph{centered-moment certificate} (Section~\ref{sec:moments}). Expanding around a center $b\in p\Zp$, we form the $n$-th centered phase moments and truncated aggregate
$$
m_n(b)=\sum_{j\in J_y}\alpha_j(z_{yj}-b)^n,
\qquad
T_r(b)=\sum_{n=0}^{r-1}\frac{m_n(b)}{n!}.
$$
We construct an explicit, computable remainder radius $\varepsilon_r(b)$ such that
$$
\abs{\expp(-b)A_y-T_r(b)}\le\varepsilon_r(b).
$$
Whenever $\abs{T_r(b)}>\varepsilon_r(b)$, unit-invariance $\abs{\expp(-b)}=1$ immediately certifies $\abs{A_y}=\abs{T_r(b)}$. We prove that the remainder bound $\varepsilon_r(b)$ is globally minimized over $b\in p\Zp$ at one of the active phases $z_{yj_0}$ (Proposition~\ref{prop:center-choice}), and that for any nonzero aggregate the certificate succeeds at sufficiently high order $r$ (Proposition~\ref{prop:moment-eventual}). For rational inputs, this provides an entirely exponential-free rational certification algorithm.

\subsection*{A certified inference layer for \texorpdfstring{$p$-adic}{p-adic} learning}

In \emph{Learning with the $p$-adics} \cite{Martins}, multiclass classifiers assign to each class $k\in K$ a linear or kernel discriminant $f_k(x)=w_k^\top x+b_k\in\Qp$ and decode via
$$
\hat y(x)=\argmin_{k\in K} \abs{f_k(x)}.
$$
Although setting $A_k=1/f_k$ formally maps this rule into argmax readout, one computes verified approximations of the discriminants $f_k$ rather than their reciprocals. Crucially, the roles of resolved and unresolved coordinates are inverted: the winning class has the \emph{smallest} norm (largest valuation) and is therefore the \emph{last} coordinate to be resolved as precision increases.

In Section~\ref{sec:discriminants}, we establish a necessary and sufficient condition for verified discriminant data to uniquely determine the exact minimizer set $M^*(f)$, ties included (Theorem~\ref{thm:argmin}). At common integer precision $p^{-N}$, Corollary~\ref{cor:argmin-precision} proves that the sharp precision threshold is
$$
N \ge v^{(2)} + 1,
$$
where $v^{(2)}$ is the \emph{second-largest} valuation among $\{v_p(f_k)\}_{k\in K}$. Consequently, when a unique winner exists ($v^{(1)}>v^{(2)}$), the optimal class is rigorously certified at precision $v^{(2)}+1$ even while the winning discriminant $f_{k^*}$ itself remains unresolved (indistinguishable from zero at precision $p^{-N}$). This provides a complete, certified finite-precision inference layer for $p$-adic classification.

\subsection*{Application to \texorpdfstring{$p$-adic}{p-adic} path-integral actions}

In Section~\ref{sec:affine}, we apply our certification machinery to discrete action functionals of the form recently introduced by Hu and Kim \cite{HuKim2026} for $p$-adic path integrals, evaluated along $k$-step uniform affine chains. We uncover two structural phenomena:
\begin{enumerate}
\item \emph{The action center:} For free propagation $z(y,x)=q(y-x)^2$ with balanced coefficients ($\sum_j\alpha_j=0$) and $\sigma_1=\sum_j\alpha_j x_j\ne 0$, the first centered moment factors affinely as $\Phi_y = -2q\sigma_1(y-c_*)$, where $c_*=(\sum_j\alpha_j x_j^2)/(2\sigma_1)$ is an explicit \emph{action center}. Outside a certified exceptional ball around $c_*$, the amplitude norm is exactly proportional to the ultrametric distance $\abs{y-c_*}$ (Theorem~\ref{thm:center}). For two balanced sources, this exceptional ball vanishes identically (Corollary~\ref{cor:two-free}).
\item \emph{Zero effective coupling degeneracy:} In the presence of a quadratic potential $V(X)=v_0+v_1X+v_2X^2$, we identify a critical parameter locus $\lambda=0$ at which the phase $z(y,x)$ separates additively into $F(y)+G(x)$. At this degeneracy, the unmasked readout becomes exactly output-independent regardless of coefficient balancing (Corollary~\ref{cor:degenerate}).
\end{enumerate}

\subsection*{Summary of contributions and organization}

Positioned within the numerical analysis literature on $p$-adic precision tracking \cite{CRV,Car} and verified scientific computing \cite{Rum,BHH}, our main contributions are: (i) readout and discriminant certificates with exact tie handling and sharp integer precision thresholds ($v_*+1$ for amplitudes, $v^{(2)}+1$ for discriminants), plus an explicit TV error ledger (Section~\ref{sec:readout}); (ii) summand-independent perturbation bounds and exponential-free centered-moment certificates with optimal center selection and a complexity crossover analysis (Sections~\ref{sec:exp}--\ref{sec:moments}); (iii) the analytical structure of affine-orbit kernels---action centers, midpoint confinement, and exact zero-coupling factorization (Section~\ref{sec:affine}); and (iv) adaptive/hybrid refinement algorithms and benchmark experiments (Sections~\ref{sec:adaptive}--\ref{sec:implementation}).

%=====================================================================
\section{Preliminaries}\label{sec:prelim}
%=====================================================================

\subsection{Verified representatives}

For nonzero $x\in\Qp$, let $v_p(x)\in\Z$ denote its $p$-adic valuation, $\abs{x}=p^{-v_p(x)}$, and set $v_p(0)=+\infty$, $\abs{0}=0$. The ultrametric (strong triangle) inequality states that
\begin{equation}\label{eq:ultra}
\abs{x+y}\le\max\{\abs{x},\abs{y}\},
\end{equation}
with equality whenever $\abs{x}\ne\abs{y}$.
For $a\in\Qp$ and $r\ge0$, we write $B(a,r)=\{x\in\Qp:\abs{x-a}\le r\}$ for the closed ball of radius $r$.

\begin{definition}[Verified representative]\label{def:verified}
A \emph{verified representative} of $A\in\Qp$ is a pair $(\tA,\delta)\in\Qp\times\R_{\ge0}$ such that $\abs{A-\tA}\le\delta$.
A zero radius ($\delta=0$) denotes exact knowledge ($A=\tA$).
\end{definition}

In verified computation, the error radius $\delta$ is a mathematically guaranteed bound derived from truncation and propagation theorems, rather than an empirical estimate obtained by comparing two finite-precision runs. For $A\in\Zp$, exact knowledge of $A\bmod p^N$ supplies a verified representative with radius $\delta=p^{-N}$.

\subsection{The \texorpdfstring{$p$-adic}{p-adic} exponential}

For an odd prime $p$, the $p$-adic exponential series $\expp(z)=\sum_{n=0}^\infty z^n/n!$ converges on the disk $p\Zp=\{z\in\Qp:\abs{z}\le p^{-1}\}$ and satisfies the fundamental identities \cite{Sch,Rob}:
\begin{align}
\expp(z+w)&=\expp(z)\expp(w),\label{eq:add}\\
\abs{\expp(z)}&=1,\label{eq:unit}\\
\abs{\expp(z)-1}&=\abs{z},\label{eq:minusone}\\
\abs{\expp(z)-\expp(w)}&=\abs{z-w}
\qquad(z,w\in p\Zp).\label{eq:isometry}
\end{align}

\begin{remark}[Extension to $p=2$ and choice of character]\label{rem:character}
The readout certificates of Section~\ref{sec:readout} rely solely on \eqref{eq:ultra} and hold verbatim for $p=2$. The exponential aggregation results are stated for odd $p$; for $p=2$ the convergence and isometric domain of $\exp_2$ is $4\Z_2$. The isometry \eqref{eq:isometry} is strictly stronger than unit-valuedness: an additive character taking values in the primitive $p$-th roots of unity $\zeta\in\mathbb C_p$ satisfies $\abs{\zeta}=1$ yet $\abs{\zeta-1}=p^{-1/(p-1)}$. Our exponential bounds therefore specifically exploit the isometric property of $\expp$ on $p\Zp$.
\end{remark}

\subsection{Readouts and tie-breaking}

\begin{definition}\label{def:readout}
For $A\in\Qp^Y$ not identically zero, $\beta>0$, and prior weights $h_y>0$, define the probability distribution $P_{\beta,h}(\cdot\mid A)$ by \eqref{eq:readout-intro}, and write $P_\beta$ when $h\equiv1$.
The \emph{maximizer set} is $\argmax_{y\in Y} h_y\abs{A_y}^\beta$, and the \emph{top-ranked output} is its first element with respect to the fixed total ordering of $Y$.
\end{definition}

If $A_y=0$ for all $y\in Y$, the readout \eqref{eq:readout-intro} is undefined; a sound certification procedure must distinguish this exact vanishing from both a uniform distribution and an unresolved finite-precision state. For probability vectors $P,Q$ on $Y$, the total variation distance is $\TV(P,Q)=\frac12\sum_{y\in Y}\rabs{P(y)-Q(y)}$. We adopt the convention $\max\varnothing=0$ for maxima of nonnegative quantities.

%=====================================================================
\section{Readout and discriminant certificates}\label{sec:readout}
%=====================================================================

\subsection{Resolved absolute values and geometric interpretation}

\begin{lemma}[Ultrametric norm stability]\label{lem:stability}
Suppose $\abs{A-\tA}\le\delta$ with $\delta\ge0$. Then
$$
\abs{A}>\delta
\quad\Longleftrightarrow\quad
\abs{\tA}>\delta.
$$
In this case $\abs{A}=\abs{\tA}$; otherwise both $\abs{A}\le\delta$ and $\abs{\tA}\le\delta$.
\end{lemma}

\begin{proof}
If $\abs{\tA}>\delta$, then $\abs{A-\tA}\le\delta<\abs{\tA}$, so the equality case of \eqref{eq:ultra} applied to $A=\tA+(A-\tA)$ gives $\abs{A}=\max\{\abs{\tA},\abs{A-\tA}\}=\abs{\tA}>\delta$. Conversely, if $\abs{A}>\delta$, the same argument with the roles of $A$ and $\tA$ interchanged (valid because $\abs{\tA-A}\le\delta$ as well) gives $\abs{\tA}=\abs{A}>\delta$. The complementary case is exactly $\abs{A}\le\delta$ and $\abs{\tA}\le\delta$.
\end{proof}

Given verified representatives $(\tA_y,\delta_y)$ for $y\in Y$, we define the \emph{resolved set} $I$ and \emph{unresolved set} $U$ by
$$
I=\{y\in Y:\abs{\tA_y}>\delta_y\},
\qquad U=Y\setminus I.
$$
By Lemma~\ref{lem:stability}, $I=\{y\in Y:\abs{A_y}>\delta_y\}$. This encodes a geometric duality: in amplitude maximization the largest norms resolve first, whereas in discriminant minimization (Section~\ref{sec:discriminants}) the smallest norm resolves last.

\subsection{A common error radius}\label{subsec:common}

\begin{theorem}\label{thm:common}
Suppose $\max_{y\in Y}\abs{A_y-\tA_y}\le\delta$, $\beta>0$, and
$$
I=\{y\in Y:\abs{\tA_y}>\delta\}\ne\varnothing.
$$
Put
$$
S_I=\sum_{y\in I}\abs{\tA_y}^{\beta},
\qquad B_U=\vert{}U\vert{}\delta^\beta.
$$
Then:
\begin{enumerate}
\item $\abs{A_y}=\abs{\tA_y}$ for every $y\in I$;
\item every exact maximizer of $\abs{A_y}$ lies in $I$, and the complete
maximizer set $\argmax_{y\in Y}\abs{A_y}$ is determined exactly by the representatives on $I$;
\item the conditional probability vector
$$
Q(y)=
\begin{cases}
\abs{\tA_y}^{\beta}/S_I,&y\in I,\\
0,&y\in U
\end{cases}
$$
satisfies the exact identity and upper bound
\begin{equation}\label{eq:common-tv}
\TV(P_\beta,Q)=P_\beta(U)
\le\frac{B_U}{S_I+B_U}.
\end{equation}
\end{enumerate}
\end{theorem}

\begin{proof}
Assertion (1) is Lemma~\ref{lem:stability}. Since $\abs{A_y}>\delta$ on $I\ne\varnothing$ and $\abs{A_y}\le\delta$ on $U$, every maximizer must belong to $I$, proving (2). For (3), let $T=\sum_{y\in U}\abs{A_y}^{\beta}\le B_U$.
Then $Q=P_\beta(\cdot\mid I)$ and
$P_\beta(U)=T/(S_I+T)$. On $I$, $Q(y)-P_\beta(y)=Q(y)P_\beta(U)\ge 0$, and on $U$, $Q(y)=0$.
Thus the $\ell^1$ distance between $P_\beta$ and $Q$ is $2P_\beta(U)$. Monotonicity of $t\mapsto t/(S_I+t)$ yields the upper bound in \eqref{eq:common-tv}.
\end{proof}

\begin{remark}\label{rem:zero}
If $I=\varnothing$, the test at radius $\delta$ has not resolved a nonzero amplitude.
It does not follow that $A\equiv 0$. However, exact zero representatives with zero radii ($\tA_y=0$, $\delta_y=0$) do certify vanishing; if this holds for every $y\in Y$, the readout is certified to be undefined.
\end{remark}

\subsection{Different radii and prior weights}

When outputs are evaluated by adaptive refinement or centered moments, their verified radii $\delta_y$ generally differ.

\begin{theorem}\label{thm:weighted}
Let $\abs{A_y-\tA_y}\le\delta_y$, $\beta>0$, and $h_y>0$, with
$I=\{y\in Y:\abs{\tA_y}>\delta_y\}\ne\varnothing$. Put
$$
w_y=h_y\abs{\tA_y}^{\beta}\quad(y\in I),\qquad
S_I=\sum_{y\in I}w_y,\qquad
B_U=\sum_{y\in U}h_y\delta_y^\beta.
$$
Then $Q(y)=w_y/S_I$ on $I$, and $Q(y)=0$ on $U$, satisfies
\begin{equation}\label{eq:weighted-tv}
\TV(P_{\beta,h},Q)=P_{\beta,h}(U)
\le\frac{B_U}{S_I+B_U}.
\end{equation}
Let $W=\max_{y\in I}w_y$. If the dominance condition
\begin{equation}\label{eq:rank}
W>\max_{y\in U}h_y\delta_y^\beta
\end{equation}
holds, then the exact maximizer set is
$\{y\in I:w_y=W\}$, and its first element is the exact top-ranked output.
\end{theorem}

\begin{proof}
For every $y\in U$, Lemma~\ref{lem:stability} gives $\abs{A_y}\le\delta_y$, so $h_y\abs{A_y}^{\beta}\le h_y\delta_y^\beta$.
The conditioning proof of Theorem~\ref{thm:common} applies directly to the weighted sums to give \eqref{eq:weighted-tv}.
Under \eqref{eq:rank}, every unresolved weight is strictly below $W$, so no element of $U$ can attain or exceed the maximum resolved weight $W$.
\end{proof}

\begin{remark}[Norm certificates suffice]\label{rem:norm-cert}
The conclusions of Theorem~\ref{thm:weighted} (and, with $h\equiv1$, of Theorem~\ref{thm:common}) depend on the verified data only through the two norm-level facts $\abs{A_y}=\abs{\tA_y}$ on $I$ and $\abs{A_y}\le\delta_y$ on $U$. We therefore call any pair $(\abs{\tA_y},\delta_y)$ satisfying these two facts a \emph{norm certificate} for $A_y$. Every verified representative induces a norm certificate, but the converse fails: a norm certificate need not identify a $p$-adic value $\tA_y$ with $\abs{A_y-\tA_y}\le\delta_y$. The centered-moment certificate of Section~\ref{sec:moments} is the key example of a norm certificate that is not a value representative, and it may be substituted verbatim into Theorem~\ref{thm:weighted}, Theorem~\ref{thm:common}, and the error ledger of Proposition~\ref{prop:real}.
\end{remark}

\begin{example}[Necessity of dominance under heterogeneous radii]\label{ex:radii}
Let $p=3$ and $h\equiv 1$. Represent $A_1=9$ by $(9,3^{-3})$ and $A_2=3$ by
$(0,3^{-1})$. Then coordinate $1$ is resolved ($\abs{9}=3^{-2}>3^{-3}$) and coordinate $2$ is unresolved ($\abs{0}\le 3^{-1}$), yet
$\abs{A_2}=3^{-1}>3^{-2}=\abs{A_1}$. Thus $I\ne\varnothing$ alone does not certify the maximizer when the error radii differ; condition \eqref{eq:rank} correctly detects the ambiguity since $W=3^{-2\beta}\not> 3^{-\beta}$.
\end{example}

\subsection{Sharp integer absolute precision threshold}\label{subsec:integer}

\begin{proposition}\label{prop:sharp}
Suppose output $y$ is known with error radius $\delta_y=p^{-N_y}$ for $N_y\in\Z$. Then
$$
y\in I\quad\Longleftrightarrow\quad N_y>v_p(A_y).
$$
For a nonzero amplitude family, put $v_*=\min_{y\in Y}v_p(A_y)$.
At a common integer absolute precision $N$,
$$
I\ne\varnothing\quad\Longleftrightarrow\quad N\ge v_*+1.
$$
Thus $v_*+1$ is the sharp least integer absolute precision at which
Theorem~\ref{thm:common} certifies the exact maximizer set and top-ranked output. If an evaluator imposes a floor $N\ge N_{\min}$, the least sufficient precision is $\max\{N_{\min},v_*+1\}$.
\end{proposition}

\begin{proof}
By Lemma~\ref{lem:stability}, $y\in I$ is equivalent to $\abs{A_y}>p^{-N_y}$, which in turn is equivalent to $v_p(A_y)<N_y$, i.e., $N_y\ge v_p(A_y)+1$. Taking the minimum over $y\in Y$ yields the threshold $v_*+1$.
\end{proof}

Although $v_*$ is unknown a priori, Proposition~\ref{prop:sharp} guarantees that any iterative schedule of increasing integer precisions will succeed at the very first visited precision $N\ge v_*+1$, and that no lower common integer precision can ever yield a nonempty resolved set.

\subsection{The error ledger: Real-valued output and dyadic rounding}

In practice, the rational conditional distribution $Q$ (when $\beta\in\N$ and $h_y\in\Q_{>0}$) is often converted into a fixed-point or floating-point vector $\widehat Q$ for downstream consumption.

\begin{proposition}[Two-part error ledger]\label{prop:real}
Under the hypotheses of Theorem~\ref{thm:weighted}, suppose $\widehat Q$ is a probability
vector satisfying $\TV(\widehat Q,Q)\le\varepsilon_{\rm real}$. Then
$$
\TV(\widehat Q,P_{\beta,h})
\le
\min\left\{1,\frac{B_U}{S_I+B_U}+\varepsilon_{\rm real}\right\}.
$$
\end{proposition}

\begin{proof}
Combine \eqref{eq:weighted-tv} with the triangle inequality $\TV(\widehat Q,P_{\beta,h})\le\TV(P_{\beta,h},Q)+\TV(Q,\widehat Q)$ and the trivial bound $\TV\le 1$.
\end{proof}

\begin{proposition}[Normalized dyadic rounding]\label{prop:dyadic}
Let $Q$ be a rational probability vector on $Y$ and let $b\in\N$ be a bit budget.
There exist nonnegative integers $(n_y)_{y\in Y}$ with $\sum_{y\in Y} n_y=2^b$ such that
$\widehat Q(y)=n_y/2^b$ satisfies
$$
\max_{y\in Y}\rabs{Q(y)-\widehat Q(y)}\le2^{-b},
\qquad
\TV(Q,\widehat Q)\le\frac{\vert{}Y\vert{}}{2^{b+1}}.
$$
\end{proposition}

\begin{proof}
Write $2^bQ(y)=a_y+r_y$ with $a_y=\lfloor 2^bQ(y)\rfloor\in\Nzero$ and
$0\le r_y<1$. The deficit
$L=2^b-\sum_{y\in Y} a_y=\sum_{y\in Y} r_y$ is an integer satisfying $0\le L<\vert{}Y\vert{}$.
Increment $a_y$ by $1$ at the $L$ coordinates with the largest remainders $r_y$ (breaking ties by the fixed ordering of $Y$). Then $\vert{}Q(y)-\widehat Q(y)\vert{}\le 2^{-b}$ for every $y$, and summing $\frac12\vert{}Q(y)-\widehat Q(y)\vert{}$ gives $\vert{}Y\vert{}/2^{b+1}$.
\end{proof}

\begin{example}[Output rounding can invert the discrete ranking]\label{ex:rounding}
Let $p=3$, $A=(3,1)$, $h\equiv 1$, and $\beta=1$. The exact readout is
$Q=(1/4,3/4)$, whose unique maximizer is the second coordinate.
At dyadic precision $b=1$, Proposition~\ref{prop:dyadic} rounds $Q$ to $\widehat Q=(1/2,1/2)$.
Naively applying tie-breaking to $\widehat Q$ selects the \emph{first} coordinate.
\end{example}

Example~\ref{ex:rounding} underscores a vital architectural principle for numerical software: an implementation must report the certified maximizer set (determined from exact $p$-adic valuations on $I$) and the rounded probability vector $\widehat Q$ as separate fields in the certificate record.

\subsection{Discriminant readouts for \texorpdfstring{$p$-adic}{p-adic} classification}\label{sec:discriminants}

Let $K$ be a finite nonempty class set with a fixed total ordering,
and let $f=(f_k)_{k\in K}\in\Qp^K$ be the class discriminants of a $p$-adic classifier \cite{Martins}. The hard classification decision is the minimizer set
$$
M^*(f)=\argmin_{k\in K}\abs{f_k}.
$$
Suppose verified representatives $(\tf_k,\delta_k)$ of the discriminants are given. Partition $K$ into three disjoint subsets:
$$
R=\{k\in K:\abs{\tf_k}>\delta_k\},\qquad
Z=\{k\notin R:\delta_k=0\},\qquad
U_+=\{k\notin R:\delta_k>0\},
$$
and set $m_R=\min_{k\in R}\abs{\tf_k}$ (with $m_R=+\infty$ if $R=\varnothing$). By Lemma~\ref{lem:stability},
$\abs{f_k}=\abs{\tf_k}>0$ on $R$, $f_k=\tf_k=0$ on $Z$, and
$\abs{f_k}\le\delta_k$ on $U_+$. We say that the verified data \emph{determine $M^*$} if $M^*(f')=M^*(f)$ for every $f'\in\Qp^K$ satisfying
$\abs{f'_k-\tf_k}\le\delta_k$ for all $k\in K$.

\begin{theorem}[Necessary and sufficient condition for certified argmin]\label{thm:argmin}
The verified discriminant data $(\tf_k,\delta_k)_{k\in K}$ determine $M^*$ if and only if one of the following mutually exclusive conditions holds:
\begin{enumerate}
\item $U_+=\varnothing$. Then $M^*=Z$ if $Z\ne\varnothing$, and
$M^*=\{k\in R:\abs{\tf_k}=m_R\}$ otherwise.
\item $Z=\varnothing$, $U_+=\{j\}$ is a singleton, and $\delta_j<m_R$. Then
$M^*=\{j\}$.
\end{enumerate}
\end{theorem}

\begin{proof}
For each $j\in U_+$ we have $\abs{\tf_j}\le\delta_j$, so by ultrametricity
$B(\tf_j,\delta_j)=B(0,\delta_j)$. This ball contains $0$ as well as every power
$p^n$ satisfying $p^{-n}\le\delta_j$.

\emph{Sufficiency.} In case (1), $K=R\cup Z$; every admissible $f'$ vanishes identically on $Z$ and has exact norms $\abs{f'_k}=\abs{\tf_k}>0$ on $R$, fixing $M^*(f')$. In case (2), $K=R\cup\{j\}$, and every admissible $f'$ satisfies
$\abs{f'_j}\le\delta_j<m_R=\min_{k\in R}\abs{f'_k}$, so $M^*(f')=\{j\}$.

\emph{Necessity.} Suppose neither (1) nor (2) holds, so $U_+\ne\varnothing$.
In each case below we keep $f'_k=f''_k=f_k$ for $k\notin U_+$, and
fix an integer $n$ with $p^{-n}\le\min_{l\in U_+}\delta_l$.
\begin{itemize}
\item If $\vert{}U_+\vert{}\ge 2$ with distinct $j,j'\in U_+$, set $f'_j=0$ and
$f'_l=p^n$ for $l\in U_+\setminus\{j\}$, and define $f''$ analogously with $j'$ in place of $j$. Then
$M^*(f')=Z\cup\{j\}\ne Z\cup\{j'\}=M^*(f'')$.
\item If $U_+=\{j\}$ and $Z\ne\varnothing$, set $f'_j=0$ and
$f''_j=p^n$. Then $M^*(f')=Z\cup\{j\}\ne Z=M^*(f'')$.
\item If $U_+=\{j\}$, $Z=\varnothing$, and $\delta_j\ge m_R$, then
$R\ne\varnothing$ and $m_R=p^{-v}$ for some $v\in\Z$. Set $f'_j=0$
and $f''_j=p^v\in B(0,\delta_j)$. Then
$M^*(f')=\{j\}$, whereas $M^*(f'')=\{j\}\cup\{k\in R:\abs{\tf_k}=m_R\}\supsetneq\{j\}$.
\end{itemize}
In all three cases, $f'$ and $f''$ are consistent with the verified data but yield distinct minimizer sets.
\end{proof}

\begin{corollary}[Sharp integer precision threshold for argmin decoding]
\label{cor:argmin-precision}
Suppose every discriminant $f_k$ is evaluated at a common positive error radius $p^{-N}$ ($N\in\Z$). Let
$v^{(1)}\ge v^{(2)}$ be the two largest elements, counted with multiplicity, of the valuation multiset
$\{v_p(f_k):k\in K\}$, with $v_p(0)=+\infty$ (and $v^{(2)}=-\infty$ if $\vert{}K\vert{}=1$). If at least two discriminants vanish identically, then $v^{(1)}=v^{(2)}=+\infty$; the threshold below is then $+\infty$, reflecting that no finite common precision can distinguish such a zero tie. Then the data determine $M^*$ if and only if
$$
N\ge v^{(2)}+1.
$$
In particular, when a unique minimizer exists ($v^{(1)}>v^{(2)}$), the decision is certified at precision $N=v^{(2)}+1$ even though the winning discriminant itself remains unresolved whenever $N\le v^{(1)}$.
\end{corollary}

\begin{proof}
Since $p^{-N}>0$, we have $Z=\varnothing$, and by Proposition~\ref{prop:sharp} applied to $f$,
$U_+=\{k\in K:v_p(f_k)\ge N\}$. Every resolved norm on $R$ is at least $p^{-(N-1)}>p^{-N}$, so the strict inequality
$\delta_j<m_R$ holds automatically whenever $U_+=\{j\}$. By Theorem~\ref{thm:argmin}, $M^*$ is determined if and only
if $\vert{}U_+\vert{}\le1$, which is equivalent to $N>v^{(2)}$, i.e., $N\ge v^{(2)}+1$.
\end{proof}

\begin{example}\label{ex:disc}
Let $p=3$, $K=\{0,1,2\}$, and $f=(1,3,9)$, so
$v^{(1)}=2$ and $v^{(2)}=1$. At $N=1$ ($\delta=3^{-1}$), the modular representatives are
$(1,0,0)$, giving $R=\{0\}$ and $U_+=\{1,2\}$; the winner between classes $1$ and $2$ is unresolved. At $N=2=v^{(2)}+1$ ($\delta=3^{-2}$), the representatives are $(1,3,0)$, giving $R=\{0,1\}$,
$U_+=\{2\}$, and $m_R=3^{-1}>3^{-2}$. Theorem~\ref{thm:argmin}(2) certifies $M^*=\{2\}$ without needing $N=3$ to resolve $f_2=9$.
\end{example}

\begin{proposition}[Soft discriminant readout bound]\label{prop:soft-disc}
Assume $f_k\ne0$ for all $k\in K$, $\beta>0$, and $h_k>0$. Let
$P=P_{\beta,h}(\cdot\mid A)$ with $A_k=1/f_k$, and let
$U=K\setminus R$. Put
$$
S_R=\sum_{k\in R}h_k\abs{\tf_k}^{-\beta},
\qquad
L_U=\sum_{j\in U}h_j\delta_j^{-\beta}.
$$
Then the total probability mass on the suboptimal resolved set $R$ satisfies
$$
P(R)\le\frac{S_R}{S_R+L_U}.
$$
If $U=\varnothing$, then $P(k)=h_k\abs{\tf_k}^{-\beta}/S_R$ exactly.
If $U=\{j\}$ is a singleton, then against the point mass $\mathbf 1_{\{j\}}$ at the certified winner $j$,
$$
\TV(P,\mathbf 1_{\{j\}})=P(R)
\le\frac{S_R}{S_R+h_j\delta_j^{-\beta}}.
$$
\end{proposition}

\begin{proof}
Since $f_k\ne0$, every $j\in U$ has $\delta_j>0$ and
$0<\abs{f_j}\le\delta_j$, whence $h_j\abs{A_j}^{\beta}\ge
h_j\delta_j^{-\beta}$. On $R$, $\abs{A_k}=\abs{\tf_k}^{-1}$ exactly.
Thus $P(R)=S_R/(S_R+T)$ where $T=\sum_{j\in U}h_j\abs{A_j}^\beta\ge L_U$. Monotonicity of $t\mapsto S_R/(S_R+t)$ yields the bound, and for $U=\{j\}$ we have $\TV(P,\mathbf 1_{\{j\}})=1-P(j)=P(R)$.
\end{proof}

%=====================================================================
\section{Verified exponential aggregation}\label{sec:exp}
%=====================================================================

\subsection{Taylor remainder bounds}

For $r\in\N$ and $s\in\N$, define the valuation exponent
\begin{equation}\label{eq:rho}
\rho_r(s)=
\left\lceil rs-\frac{r-1}{p-1}\right\rceil,
\end{equation}
and let
$R_r(z)=\expp(z)-\sum_{n=0}^{r-1}z^n/n!$ denote the $r$-th Taylor remainder.

\begin{lemma}\label{lem:tail}
For nonzero $z\in p\Zp$ with $s=v_p(z)\ge 1$,
$$
\abs{R_r(z)}\le p^{-\rho_r(s)}.
$$
In the low-order regime $1\le r\le p-1$, this simplifies to
$\abs{R_r(z)}\le\abs{z}^{r}$.
Furthermore, for any target integer precision $N\ge1$, choosing the truncation order
\begin{equation}\label{eq:cutoff}
L\ge
\max\left\{
1,\
\left\lceil
\frac{(p-1)(N-1)}{(p-1)s-1}
\right\rceil
\right\}
\end{equation}
guarantees $\abs{R_L(z)}\le p^{-N}$.
\end{lemma}

\begin{proof}
By Legendre's formula, the $p$-adic valuation of $n!$ satisfies
$$
v_p(n!)=\frac{n-s_p(n)}{p-1}\le\frac{n-1}{p-1}
\qquad(n\ge1),
$$
where $s_p(n)\ge 1$ is the sum of the base-$p$ digits of $n$.
Consequently,
$$
v_p(z^n/n!)=ns-v_p(n!)\ge ns-\frac{n-1}{p-1}.
$$
Because $s\ge 1$ and $p\ge 3$, the right-hand side is strictly increasing in $n$. Since $v_p(z^n/n!)$ is an integer, every term with $n\ge r$ has valuation at least $\rho_r(s)$, and the ultrametric inequality \eqref{eq:ultra} extends this bound to the convergent series $R_r(z)$.
For $1\le r\le p-1$, $0\le (r-1)/(p-1)<1$, so $\rho_r(s)=rs$.
Finally, setting $d=(p-1)s-1\ge 1$, we have $\rho_L(s)=\lceil (dL+1)/(p-1)\rceil$, so $\rho_L(s)\ge N$ holds if and only if $dL+1 > (p-1)(N-1)$, which is equivalent to $dL\ge(p-1)(N-1)$.
\end{proof}

\subsection{Aggregate perturbations}

To provide a rigorous error ledger for the aggregate $A_y = \sum_{j \in J_y} \alpha_j \expp(z_{yj})$, we decompose the computation into three distinct sources of error. Let $\widetilde \alpha_j$, $\widetilde z_{yj}$, and $\widetilde E_{yj}$ denote the finite-precision coefficient, approximate phase, and modular exponential result computed by our software. We assume the following verified error bounds:

\begin{enumerate}
    \item \textbf{Coefficient Error:} The approximation of the coefficient $\alpha_j$ is controlled by $\abs{\alpha_j-\widetilde\alpha_j}\le e_{\alpha,j}$.
    \item \textbf{Phase Error and Admissibility:} The phases $z_{yj}$ and $\widetilde z_{yj}$ are guaranteed to lie within the convergence disk $p\Zp$, ensuring that the exponential series converges and the isometry \eqref{eq:isometry} holds, with $\abs{z_{yj}-\widetilde z_{yj}}\le e_{z,yj}$.
    \item \textbf{Truncation and Rounding Error:} The computed value $\widetilde E_{yj}$ approximates the true exponential $\expp(\widetilde z_{yj})$ with radius $e_{E,yj}$, satisfying $\abs{\expp(\widetilde z_{yj})-\widetilde E_{yj}}\le e_{E,yj} \le 1$.
\end{enumerate}

The condition $e_{E,yj} \le 1$ is essential for the \emph{ultrametric confinement} of the error. Specifically, since $\abs{\expp(\widetilde z_{yj})} = 1$ for all $\widetilde z_{yj} \in p\Zp$, the strong triangle inequality \eqref{eq:ultra} implies:
\[
\begin{aligned}
\abs{\widetilde E_{yj}}
&= \abs{\expp(\widetilde z_{yj}) - (\expp(\widetilde z_{yj}) - \widetilde E_{yj})} \\
&\le \max\{\abs{\expp(\widetilde z_{yj})}, \abs{\expp(\widetilde z_{yj}) - \widetilde E_{yj}}\}
\le \max\{1, e_{E,yj}\} = 1.
\end{aligned}
\]
This confirms that the numerical approximation $\widetilde E_{yj}$ resides in the unit disk $\Zp$. Consequently, when we decompose the error of a single source $j \in J_y$ into the identity
\[
\begin{aligned}
\alpha_j\expp(z_{yj}) - \widetilde\alpha_j\widetilde E_{yj}
={}& \alpha_j\bigl(\expp(z_{yj})-\expp(\widetilde z_{yj})\bigr)
+ \alpha_j\bigl(\expp(\widetilde z_{yj})-\widetilde E_{yj}\bigr) \\
&+ (\alpha_j-\widetilde\alpha_j)\widetilde E_{yj},
\end{aligned}
\]
we apply the isometry \eqref{eq:isometry} to the first term and the property $\abs{\widetilde E_{yj}}\le 1$ to the third term to obtain a tight error bound for each source $j$.

\begin{theorem}[Summand-independent aggregate perturbation]\label{thm:aggregate}
The error of the approximate aggregate $\tA_y = \sum_{j \in J_y} \widetilde\alpha_j \widetilde E_{yj}$ satisfies
\[
\abs{A_y - \tA_y} \le \max_{j \in J_y} \max\bigl\{ e_{\alpha,j}, \ \abs{\alpha_j}e_{z,yj}, \ \abs{\alpha_j}e_{E,yj} \bigr\}.
\]
An empty active set $J_y = \varnothing$ yields the exact representative $(0,0)$.
\end{theorem}

\begin{proof}
By the ultrametric inequality, the absolute value of the sum is bounded by the maximum absolute value of the individual source errors derived above. Unlike standard archimedean summation, where $\abs{\sum \delta_j} \le \sum \abs{\delta_j}$ forces a factor of $|J_y|$, the $p$-adic strong triangle inequality $\abs{\sum \delta_j} \le \max \abs{\delta_j}$ renders the total aggregate error independent of the number of sources.
\end{proof}

In sharp contrast to real floating-point summation---where worst-case error bounds grow linearly with $\vert{}J_y\vert{}$ (or as $\sqrt{\vert{}J_y\vert{}}$ under probabilistic models \cite{Hig})---Theorem~\ref{thm:aggregate} carries no summand-count factor: increasing the number of sources increases arithmetic work but never inflates the $p$-adic error radius.

\begin{corollary}\label{cor:modular}
Suppose $\alpha_j\in\Q\cap\Zp$ and all active phases $z_{yj}$ are exact rationals
in $p\Zp$. Evaluating coefficients, Taylor-truncated exponentials (with cutoff \eqref{eq:cutoff}), and their sum
modulo $p^N$ produces a verified representative $\tA_y$ with
$\abs{A_y-\tA_y}\le p^{-N}$.
\end{corollary}

\subsection{Phase shifts and two-source aggregates}

A phase shift by $b_y\in p\Zp$ multiplies $A_y$ by $\expp(b_y)$ and hence, by \eqref{eq:unit}, leaves $\abs{A_y}$ unchanged; this unit-shift invariance underlies the centered-moment certificates of Section~\ref{sec:moments}. The following exact two-source identity is used repeatedly.

\begin{theorem}[Exact two-source identity]\label{thm:two}
For any $z_1,z_2\in p\Zp$ and $c\in\Qp$,
$$
\abs{c(\expp(z_1)-\expp(z_2))}
=
\abs{c}\abs{z_1-z_2}.
$$
\end{theorem}

\begin{proof}
Factor out $\expp(z_2)$ and apply \eqref{eq:add}--\eqref{eq:isometry}.
\end{proof}

\begin{example}[Nonbalanced cancellation]\label{ex:nonbalanced}
Let $p=3$, $K(y,x)=\expp(3(y-x)^2)$, sources $(x_1,x_2)=(0,1)$, and coefficients
$(\alpha_1,\alpha_2)=(1,2)$. By Lemma~\ref{lem:tail}, at $N=4$ the cutoff is $L=6$ and
$\expp(3)\equiv 1+3+\frac92+\frac92+\frac{27}{8}=\frac{131}{8}\equiv67\pmod{81}$; hence
$A_0=1+2\expp(3)\equiv 54\pmod{81}$ and $\abs{A_0}=3^{-3}$, whereas
$A_1=\expp(3)+2\equiv6\pmod9$ and $\abs{A_1}=3^{-1}$.
Thus zero-sum balancing ($\sum_j\alpha_j=0$) is not required for norm discrimination; what matters is the valuation of the leading moments, formalized next.
\end{example}

%=====================================================================
\section{Centered-moment certificates}\label{sec:moments}
%=====================================================================

Fix a finite active source set $J$, coefficients $\alpha_j\in\Qp$, and phases $z_j\in p\Zp$, and consider the aggregate
$A=\sum_{j\in J}\alpha_j\expp(z_j)$.
Put
$$
M=\max_{j\in J}\abs{\alpha_j},
\qquad
D=\max_{j,k\in J}\abs{z_j-z_k},
$$
with $M=0$ for $J=\varnothing$ and $D=0$ for $\vert{}J\vert{}\le1$.

\subsection{The centered-moment certificate}

Why expand around a center $b$ rather than evaluate each $\expp(z_j)$? When $\sum_{j\in J}\alpha_j$ has high $p$-adic valuation (or vanishes), individual evaluations compute low-valuation leading digits ($1+z_j+\cdots$) that cancel only at the final summation. Summing the Taylor series around a reference phase $b\in p\Zp$ performs this cancellation \emph{exact-algebraically inside the moments}, before any truncation.

For a center $b\in p\Zp$ and order $r\in\N$, define the shifted phases, centered moments, and truncated moment aggregate by
$$
w_j=z_j-b,\qquad
m_n(b)=\sum_{j\in J}\alpha_jw_j^n,\qquad
T_r(b)=\sum_{n=0}^{r-1}\frac{m_n(b)}{n!},
$$
where $0^0=1$. Define the per-source and aggregate remainder bounds by
$$
\theta_r(w)=
\begin{cases}
p^{-\rho_r(v_p(w))},&w\ne0,\\
0,&w=0,
\end{cases}
\qquad
\varepsilon_r(b)=
\max_{j\in J}\bigl\{\abs{\alpha_j}\theta_r(w_j)\bigr\}.
$$

\begin{theorem}[Centered-moment certificate]\label{thm:moment}
For every center $b\in p\Zp$ and truncation order $r\in\N$,
\begin{equation}\label{eq:moment}
\abs{\expp(-b)A-T_r(b)}\le\varepsilon_r(b).
\end{equation}
Consequently:
$$
\abs{T_r(b)}>\varepsilon_r(b)
\quad\Longrightarrow\quad
\abs{A}=\abs{T_r(b)};
$$
otherwise $\abs{A}\le\varepsilon_r(b)$.
In the low-order regime $1\le r\le p-1$, the remainder radius takes the form
$$
\varepsilon_r(b)=
\max_{j\in J}\bigl\{\abs{\alpha_j}\abs{z_j-b}^{r}\bigr\},
$$
which is bounded above by $MD^r$ whenever $b\in\{z_j:j\in J\}$.
\end{theorem}

\begin{proof}
By \eqref{eq:add}, factoring out $\expp(b)$ yields
$$
\expp(-b)A
=
\sum_{j\in J}\alpha_j\expp(w_j)
=
T_r(b)+\sum_{j\in J}\alpha_jR_r(w_j).
$$
Applying Lemma~\ref{lem:tail} to each nonzero $w_j$ and taking the ultrametric maximum over $j\in J$ gives \eqref{eq:moment}.
Since $\abs{\expp(-b)}=1$, we have $\abs{\expp(-b)A}=\abs{A}$, and Lemma~\ref{lem:stability} yields the resolution dichotomy.
\end{proof}

When $\alpha_j,z_j,b\in\Q$, both $T_r(b)$ and $\varepsilon_r(b)$ are computed in exact rational arithmetic without evaluating a single $p$-adic exponential. By Theorem~\ref{thm:moment}, the pair $(\abs{T_r(b)},\varepsilon_r(b))$ is a norm certificate in the sense of Remark~\ref{rem:norm-cert}: it satisfies $\abs{A_y}=\abs{T_r(b)}$ whenever $\abs{T_r(b)}>\varepsilon_r(b)$, and $\abs{A_y}\le\varepsilon_r(b)$ otherwise.

\subsection{Optimal finite center selection}

A priori, choosing an optimal center $b\in p\Zp$ to minimize $\varepsilon_r(b)$ is a continuous optimization problem over an infinite compact disk. The ultrametric geometry reduces this to a finite search over the active phases themselves.

\begin{proposition}[Finite center optimality]\label{prop:center-choice}
Suppose $J\ne\varnothing$. For every $b\in p\Zp$, there exists an active source
$j_0\in J$ such that
$$
\varepsilon_r(z_{j_0})\le\varepsilon_r(b)
\qquad\hbox{simultaneously for all }r\in\N.
$$
Thus the global minimum of the remainder bound $\varepsilon_r(b)$ over $b\in p\Zp$ is attained at one of the active phases $z_j$.
\end{proposition}

\begin{proof}
Given $b\in p\Zp$, choose $j_0\in J$ minimizing $\abs{z_j-b}$. For every $j\in J$, ultrametricity gives
$$
\abs{z_j-z_{j_0}}
\le
\max\{\abs{z_j-b},\abs{z_{j_0}-b}\}
=
\abs{z_j-b}.
$$
Because $\theta_r(w)$ is a nondecreasing function of $\abs{w}$, it follows that $\abs{\alpha_j}\theta_r(z_j-z_{j_0})\le\abs{\alpha_j}\theta_r(z_j-b)$ for every $j\in J$ and $r\in\N$. Taking the maximum over $j\in J$ completes the proof.
\end{proof}

Algorithm~\ref{alg:moment} summarizes the resulting exponential-free certification procedure for a single aggregate $A_y$.

\begin{algorithm}[htbp]
\caption{Exponential-Free Centered-Moment Certificate (\texttt{CenteredMomentCert})}
\label{alg:moment}
\begin{algorithmic}[1]
\Require Prime $p\ge 3$, active sources $J\ne\varnothing$, rational coefficients $(\alpha_j)_{j\in J}$, rational phases $(z_j)_{j\in J}\subset p\Zp$, maximum moment order $r_{\max}\ge 1$.
\Ensure Certified norm $\abs{T_r}$ with radius $\varepsilon_r$ and flag $\texttt{resolved}$; when set, $\abs{A}=\abs{T_r}$ (the moment ball represents $\expp(-b)A$, whose norm equals $\abs{A}$).
\State Combine sources with identical rational phases by summing their coefficients; if all cancel, \Return $(0, 0, \texttt{True})$.
\For{$r = 1, 2, \ldots, r_{\max}$}
    \State Select $b_r \in \argmin_{b \in \{z_j : j\in J\}} \varepsilon_r(b)$ \Comment{finite center search; Proposition~\ref{prop:center-choice}}
    \State $T_r \gets \sum_{n=0}^{r-1} \frac{m_n(b_r)}{n!}$, with $m_n(b_r)=\sum_{j\in J}\alpha_j(z_j-b_r)^n \in \Q$
    \State $\varepsilon_r \gets \max_{j\in J} \bigl\{ \abs{\alpha_j} \theta_r(z_j - b_r) \bigr\}$
    \If{$\abs{T_r} > \varepsilon_r$}
        \State \Return $(\abs{T_r}, \varepsilon_r, \texttt{True})$ \Comment{Exact norm $\abs{A}=\abs{T_r}$ certified}
    \EndIf
\EndFor
\State \Return $(\abs{T_{r_{\max}}}, \varepsilon_{r_{\max}}, \texttt{False})$ \Comment{Unresolved; upper bound $\abs{A}\le\varepsilon_{r_{\max}}$ certified}
\end{algorithmic}
\end{algorithm}

\subsection{Zeroth- and first-order corollaries}

\begin{corollary}[Zeroth-order resolution]\label{cor:zeroth}
If $\abs{\sum_{j\in J}\alpha_j}>MD$ (or more generally $\abs{\sum_{j\in J}\alpha_j}>\varepsilon_1(b)$ for some $b\in p\Zp$),
then
$$
\abs{A}=\abs{\sum_{j\in J}\alpha_j}.
$$
\end{corollary}

\begin{proof}
Apply Theorem~\ref{thm:moment} with $r=1$, noting that $T_1(b)=m_0(b)=\sum_{j\in J}\alpha_j$ and $\varepsilon_1(z_{j_0})\le MD$.
\end{proof}

\begin{corollary}[Balanced first-order resolution]\label{cor:first}
Suppose $\sum_{j\in J}\alpha_j=0$, and define the linear phase moment
$\Phi=\sum_{j\in J}\alpha_jz_j$. If
$$
\abs{\Phi}>MD^2,
$$
then $\abs{A}=\abs{\Phi}$; otherwise $\abs{A}\le MD^2$.
\end{corollary}

\begin{proof}
For any phase center $b\in\{z_j:j\in J\}$, balancing gives $m_0(b)=0$ and $m_1(b)=\Phi$.
Apply Theorem~\ref{thm:moment} with $r=2\le p-1$, for which $T_2(b)=\Phi$ and $\varepsilon_2(b)\le MD^2$.
\end{proof}

\subsection{Higher order and eventual resolution}

\begin{proposition}[Single dominant moment]\label{prop:leading}
Let $1\le r\le p-1$ and choose a phase center $b\in\{z_j:j\in J\}$.
If there exists an index $0\le n_0<r$ such that
$$
\abs{m_{n_0}(b)}>\abs{m_n(b)}
\quad(\forall\, n<r,\ n\ne n_0)
\qquad\hbox{and}\qquad
\abs{m_{n_0}(b)}>MD^r,
$$
then $\abs{A}=\abs{m_{n_0}(b)}$.
\end{proposition}

\begin{proof}
For $n<p$, $v_p(n!)=0$ so $\abs{m_n(b)/n!}=\abs{m_n(b)}$. By the equality case of \eqref{eq:ultra}, the strictly dominant term $n_0$ fixes $\abs{T_r(b)}=\abs{m_{n_0}(b)}>MD^r\ge\varepsilon_r(b)$, and Theorem~\ref{thm:moment} applies.
\end{proof}

\begin{proposition}[Eventual resolution of nonzero aggregates]\label{prop:moment-eventual}
Fix any center $b\in p\Zp$. Then $\varepsilon_r(b)\to0$ as $r\to\infty$.
Consequently, if $A\ne0$, there exists a finite threshold $r_0\in\N$ such that for all $r\ge r_0$,
$$
\abs{T_r(b)}>\varepsilon_r(b)
\qquad\hbox{and}\qquad
\abs{T_r(b)}=\abs{A}.
$$
\end{proposition}

\begin{proof}
For each $j\in J$ with $w_j=z_j-b\ne 0$, we have $s_j=v_p(w_j)\ge 1$, so $\rho_r(s_j)\ge r(1-\frac{1}{p-1})\to+\infty$ as $r\to\infty$. Since $J$ is finite, $\varepsilon_r(b)\to 0$. Once $\varepsilon_r(b)<\abs{A}$, Lemma~\ref{lem:stability} applied to \eqref{eq:moment} yields $\abs{T_r(b)}=\abs{A}>\varepsilon_r(b)$.
\end{proof}

\subsection{Asymptotic complexity and crossover analysis}\label{subsec:crossover}

To compare centered moments with direct modular evaluation (Corollary~\ref{cor:modular}), fix an output $y$ with $m=\vert{}J\vert{}$ active sources and target precision $N\ge1$, and set $s_0=\min_{j\in J}v_p(z_j)\ge1$ and $s_D=\min_{j,k\in J}v_p(z_j-z_k)\ge1$, so that the phase diameter is $D=p^{-s_D}$.

\begin{proposition}[Arithmetic cost of the two certification paths]\label{prop:cost}
The modular evaluator of Corollary~\ref{cor:modular} expands each of the $m$ exponentials with
$$
L_{\rm mod}=
\max\left\{1,\left\lceil\frac{(p-1)(N-1)}{(p-1)s_0-1}\right\rceil\right\}
$$
Taylor terms, for $\mathcal{O}(m\,L_{\rm mod})$ arithmetic operations on residues modulo $p^N$, each on at most $N\log_2 p$ bits. The centered-moment test at order $r$ computes $T_r$ and $\varepsilon_r$ in $\mathcal{O}(m\,r)$ exact rational operations and certifies $\abs{A}$ whenever $\abs{T_r}>\varepsilon_r$, where
$$
\varepsilon_r\le M\,p^{-\rho_r(s_D)},\qquad
\rho_r(s_D)=\left\lceil r s_D-\frac{r-1}{p-1}\right\rceil .
$$
\end{proposition}

Thus, when algebraic cancellation resolves the aggregate at low order $r\ll L_{\rm mod}$ (e.g.\ tight phase clustering $s_D\gg s_0$, or $r\in\{1,2,3\}$), the centered-moment path uses fewer operations and no modular inversions; when $r$ or the rational bit-length grows large, the bounded-precision modular path is preferable. This motivates the hybrid policy of Section~\ref{sec:adaptive}, which caps the rational moment order at a small $r_{\max}$ before falling back to modular arithmetic.

%=====================================================================
\section{Affine-orbit kernels and \texorpdfstring{$p$-adic}{p-adic} path-integral actions}\label{sec:affine}
%=====================================================================

This section addresses a numerical question rather than a purely physical one: for which kernel structures does certified readout admit an \emph{exponential-free closed form}, allowing the exponential and moment machinery of Sections~\ref{sec:moments}--\ref{sec:adaptive} to be bypassed? The organizing fact is that the first centered moment $\Phi_y$ factors affinely. Its consequences are threefold. (i) Outside the certified exceptional ball of Theorem~\ref{thm:center}, the readout reduces to the ultrametric power law $\abs{y-c_*}^\beta$, computable in $\mathcal O(|Y|)$ exact rational operations with no $p$-adic series. (ii) Corollary~\ref{cor:two-free} closes this for two sources with no exceptional ball at all, and is the basis of the two-source speedups reported in the supplementary material. (iii) Corollary~\ref{cor:degenerate} locates the zero-coupling parameter locus at which the readout collapses to the priors exactly. Outside these regimes, the moment certificates of Section~\ref{sec:moments} bound the exceptional region that still requires series evaluation.

\subsection{The discrete affine-chain action}

Consider a $k$-step uniform affine interpolation between a source node $x\in\Qp$ and a target node $y\in\Qp$. Let $k\in\N$, $\mu,\epsilon,a\in\Qp^\times$, and let $V\in\Qp[X]$ be a polynomial potential. The chain nodes $x_i^*$ and midpoint evaluation points $q_i^*$ are given by
\begin{align}
x_i^*&=x+\frac{i}{k}(y-x)\qquad(0\le i\le k),\label{eq:orbit}\\
q_i^*&=\frac{x_i^*+x_{i+1}^*}{2}=x+\frac{2i+1}{2k}(y-x)\qquad(0\le i<k).\label{eq:midpoints}
\end{align}
Evaluating the discrete time-sliced action functional on this affine chain yields
\begin{equation}\label{eq:action}
S(y,x)=
\frac{\mu}{2k\epsilon}(y-x)^2
-\sum_{i=0}^{k-1}V(q_i^*).
\end{equation}
For admissible pairs $(y,x)$---meaning those for which the scaled phase satisfies
$z(y,x):=aS(y,x)\in p\Zp$---we define the aggregation kernel
$$
K(y,x)=\expp(z(y,x)).
$$
Because reversing $x$ and $y$ simply reverses the indexing of the midpoints $q_i^*$, both $S(y,x)=S(x,y)$ and $K(y,x)=K(x,y)$ are symmetric wherever defined.

\paragraph{Connection to Hu and Kim \cite{HuKim2026}.}
The action \eqref{eq:action} specializes the discrete time-sliced action of \cite{HuKim2026} to the uniform affine trajectory, and, when the additive character there is instantiated on $p\Zp$ via $\expp$, the kernel $K(y,x)$ is the semi-classical transition factor along the chain.

\subsection{The free action center}

When $V=0$, set $q=a\mu/(2k\epsilon)$, so that
$z(y,x)=q(y-x)^2$.
For a balanced source configuration ($\sum_{j}\alpha_j=0$), define the source moments and geometric scale parameters:
$$
\sigma_1=\sum_j\alpha_jx_j,\qquad
\sigma_2=\sum_j\alpha_jx_j^2,\qquad
M=\max_j\abs{\alpha_j},
$$
$$
\Delta=\max_{j,l}\abs{x_j-x_l},
\qquad
R_y=\max_j\abs{y-x_j},
\qquad
D_y=\max_{j,l}\abs{z(y,x_j)-z(y,x_l)}.
$$

\begin{theorem}[Free action center and exceptional ball]\label{thm:center}
Assume all active phases at output $y$ are admissible.
If $\sigma_1\ne0$, define the \emph{action center} $c_*=\sigma_2/(2\sigma_1)\in\Qp$.
Whenever $y$ lies outside the exceptional region defined by
\begin{equation}\label{eq:center-test}
\abs{q\sigma_1}\abs{y-c_*}>MD_y^2,
\end{equation}
the aggregate amplitude norm is given exactly by
\begin{equation}\label{eq:center-norm}
\abs{A_y}=\abs{q\sigma_1}\abs{y-c_*}.
\end{equation}
Moreover, $D_y\le\abs{q}\Delta R_y$, so a sufficient condition for \eqref{eq:center-norm} is
$$
\abs{\sigma_1}\abs{y-c_*}>
\abs{q}M\Delta^2R_y^2.
$$
If $\sigma_1=0$ and $\abs{q\sigma_2}>MD_y^2$, then $\abs{A_y}=\abs{q\sigma_2}$ is independent of $y$.
\end{theorem}

\begin{proof}
Expanding $(y-x_j)^2=y^2-2x_j y+x_j^2$ and using $\sum_j\alpha_j=0$ gives the first moment
$$
\Phi_y
=
\sum_j\alpha_jq(y-x_j)^2
=
q(\sigma_2-2\sigma_1y).
$$
Since $p$ is odd, $\abs{2}=1$, so $\abs{\Phi_y}=\abs{q\sigma_1}\abs{y-c_*}$ when $\sigma_1\ne 0$.
Applying Corollary~\ref{cor:first} yields \eqref{eq:center-test}--\eqref{eq:center-norm}.
Finally, factoring $z(y,x_j)-z(y,x_l)=q(x_l-x_j)\bigl((y-x_j)+(y-x_l)\bigr)$ and applying \eqref{eq:ultra} gives $D_y\le\abs{q}\Delta R_y$.
\end{proof}

When \eqref{eq:center-test} holds across all $y\in Y$, the common factor $\abs{q\sigma_1}^\beta$ cancels in \eqref{eq:readout-intro}, and the exact probability readout reduces to a pure power-law distance from the action center:
$$
P_\beta(y\mid A)
=
\frac{\abs{y-c_*}^{\beta}}
{\sum_{z\in Y}\abs{z-c_*}^{\beta}}.
$$

\begin{corollary}[Exact two-source free readout]\label{cor:two-free}
For two sources $x_1,x_2$ with balanced coefficients $(c,-c)$ and any admissible output $y$,
$$
\abs{
c\bigl(\expp(q(y-x_1)^2)-\expp(q(y-x_2)^2)\bigr)
}
=
\abs{cq(x_2-x_1)}\abs{2y-x_1-x_2}.
$$
\end{corollary}

\begin{proof}
Apply Theorem~\ref{thm:two} and factor $(y-x_1)^2-(y-x_2)^2=(x_2-x_1)(2y-x_1-x_2)$.
\end{proof}

Notably, for two balanced sources the exceptional ball around $c_*=(x_1+x_2)/2$ vanishes completely: Corollary~\ref{cor:two-free} holds at every admissible $y$ without evaluating any exponential.

\subsection{Quadratic potentials and zero effective coupling}

For a quadratic potential $V(X)=v_0+v_1X+v_2X^2$, the midpoint sums along \eqref{eq:midpoints} evaluate to
$$
\sum_{i=0}^{k-1}q_i^*=\frac{k}{2}(x+y),
\qquad
\sum_{i=0}^{k-1}(q_i^*)^2
=
kxy+\frac{4k^2-1}{12k}(y-x)^2.
$$

\begin{theorem}[Shifted action center for quadratic potentials]\label{thm:quadratic}
Let $V(X)=v_0+v_1X+v_2X^2$, and define the effective kinetic and coupling parameters
$$
Q=a\left(
\frac{\mu}{2k\epsilon}
-v_2\frac{4k^2-1}{12k}
\right),
\qquad
\lambda=a\left(
\frac{\mu}{k\epsilon}
+v_2\frac{2k^2+1}{6k}
\right).
$$
For balanced coefficients ($\sum_j\alpha_j=0$), the first moment is
$$
\Phi_y
=
-\lambda\sigma_1y
+Q\sigma_2-\frac{akv_1}{2}\sigma_1.
$$
If $\lambda\sigma_1\ne0$, define the shifted action center
$$
c_V=
\frac{Q\sigma_2-\frac{akv_1}{2}\sigma_1}
{\lambda\sigma_1}.
$$
At every output $y$ with admissible phases, $\abs{A_y}=\abs{\lambda\sigma_1}\abs{y-c_V}$ whenever $\abs{\lambda\sigma_1}\abs{y-c_V}>MD_y^2$.
\end{theorem}

\begin{proof}
Substituting the midpoint sums into \eqref{eq:action} gives
$$
z(y,x)
=
Q(y-x)^2-akv_2xy-\frac{akv_1}{2}(x+y)-akv_0.
$$
Using the identity $2Q+akv_2=\lambda$ and $\sum_j\alpha_j=0$ yields $\Phi_y$, and Corollary~\ref{cor:first} gives the norm certificate.
\end{proof}

\begin{corollary}[Exact output-independence at zero effective coupling]\label{cor:degenerate}
Consider an unmasked aggregate ($J_y=J$ and $\alpha_j$ independent of $y$) such that all phases $z(y,x_j)$ are admissible on $Y\times J$. If the effective coupling vanishes ($\lambda=0$, i.e., $\mu/\epsilon=-v_2(2k^2+1)/6$), then $\abs{A_y}$ is constant across $y\in Y$, without any balancing assumption on $(\alpha_j)$.
Consequently, if this common norm is nonzero, $P_{\beta,h}(y\mid A)=h_y/\sum_{z\in Y}h_z$.
\end{corollary}

\begin{proof}
Setting $t=akv_1/2$, we may rewrite the full quadratic phase as
$$
z(y,x)
=
\underbrace{(Qy^2-ty-akv_0)}_{F(y)}
+
\underbrace{(Qx^2-tx)}_{G(x)}
-\lambda xy.
$$
When $\lambda=0$, fix a reference output $y_0\in Y$. For every $j\in J$,
$$
z(y,x_j)-z(y_0,x_j)=F(y)-F(y_0)\in p\Zp,
$$
since both $z(y,x_j)$ and $z(y_0,x_j)$ lie in the subgroup $p\Zp$. By \eqref{eq:add},
$A_y=\expp(F(y)-F(y_0))A_{y_0}$, and $\abs{\expp(F(y)-F(y_0))}=1$ gives $\abs{A_y}=\abs{A_{y_0}}$.
\end{proof}

\subsection{Worked example: Second-order resolution at the action center}

\begin{example}[Resolving the exceptional point via $r=3$ moments]\label{ex:three}
Let $p=5$, $k=1$, $a=1$, $\mu=10$, $\epsilon=1$, $V=0$ ($q=5$), with three sources $(x_1,x_2,x_3)=(0,1,2)$, balanced coefficients $(1,1,-2)$, and output space
$Y=\{0,1,2,3,4\}$. Here
$$
\sigma_1=-3,\qquad \sigma_2=-7,\qquad c_*=\frac76\equiv2\pmod5,
\qquad \Delta=1,\qquad R_y\le1.
$$
For $y\in\{0,1,3,4\}$, $\abs{y-c_*}=1$, and Theorem~\ref{thm:center} immediately certifies $\abs{A_y}=\abs{-15}=5^{-1}$.
At the exceptional output $y=2$ ($\abs{2-c_*}=5^{-1}$), the active phases are $(20,5,0)$ and $\Phi_2=25$, so the first-order test fails with equality ($\abs{\Phi_2}=5^{-2}=MD_2^2$).
Invoking Algorithm~\ref{alg:moment} at $y=2$ with phase center $b=0$ and order $r=3$ gives
$$
m_0(0)=0,\qquad m_1(0)=25,\qquad m_2(0)=425,\qquad
T_3(0)=25+\frac{425}{2}=\frac{475}{2}.
$$
Since $\abs{T_3(0)}=5^{-2}>5^{-3}=\varepsilon_3(0)$, the third-order moment certificate resolves $\abs{A_2}=5^{-2}$ in pure rational arithmetic. For $\beta=1$, the exact readout is
$P_1=(\frac{5}{21},\frac{5}{21},\frac{1}{21},\frac{5}{21},\frac{5}{21})$, certifying the tied maximizer set $\{0,1,3,4\}$.
\end{example}

\subsection{Trajectory confinement and prefix-tree masks}

\begin{lemma}[Ultrametric trajectory confinement]\label{lem:confinement}
Let $k\ge2$ and $r>0$.
All internal chain nodes $x_i^*$ belong to $B(x,r)$ if and only if
$\abs{y-x}\le\abs{k}r$, and all midpoints $q_i^*$ belong to $B(x,r)$ if and only if
$\abs{y-x}\le\abs{2k}r$.
For odd $p$, these two conditions coincide and imply $y\in B(x,r)$.
\end{lemma}

\begin{proof}
From \eqref{eq:orbit}--\eqref{eq:midpoints}, $\abs{x_i^*-x}=(\abs{i}/\abs{k})\abs{y-x}$ and $\abs{q_i^*-x}=(\abs{2i+1}/\abs{2k})\abs{y-x}$, whose maxima over $i$ are attained at $i=1$ ($\abs{1}=1$) and $i=0$ ($\abs{1}=1$), respectively.
\end{proof}

When encoding a rooted $p$-ary tree of depth $d$ by leaves $x=\sum_{\ell=0}^{d-1}a_\ell p^\ell$, a shared prefix of length $\gamma\in\Nzero$ corresponds to the ultrametric mask $\abs{y-x}\le p^{-\gamma}$ (or $\abs{y-x}\le\abs{k}p^{-\gamma}$ when trajectory confinement within the subtree ball is enforced for $p\mid k$).

%=====================================================================
\section{Adaptive computation and hybrid certification policies}\label{sec:adaptive}
%=====================================================================

\subsection{Algorithmic formulation of the hybrid adaptive policy}

We now synthesize our readout certificates, centered moments, and modular evaluators into a complete certification algorithm (Algorithm~\ref{alg:hybrid}). Given a target total-variation tolerance $\eta\in(0,1)$ and dyadic bit precision $b$ chosen so that $\varepsilon_{\rm real}=\vert{}Y\vert{}/2^{b+1}<\eta$, the policy first attempts inexpensive centered-moment resolution up to order $r_{\max}$, and then adaptively refines only those outputs in $U$ that remain unresolved, terminating as soon as both the dominance condition \eqref{eq:rank} and the TV budget are met.

\begin{algorithm}[htbp]
\caption{Hybrid Adaptive Certified Readout (\texttt{HybridAdaptiveReadout})}
\label{alg:hybrid}
\begin{algorithmic}[1]
\Require Output space $Y$, active sets $(J_y)_{y\in Y}$, coefficients $\alpha$, phases $z$, parameters $\beta\in\N$, $h\in\Q_{>0}^Y$, TV tolerance $\eta\in(0,1)$, dyadic bits $b$ with $\varepsilon_{\rm real}=\vert{}Y\vert{}/2^{b+1}\le\eta/2$, moment cap $r_{\max}$, modular schedule $N_0, \Delta N, N_{\max}$.
\Ensure Certificate record $(\text{Maximizers}, \text{TopOutput}, \widehat Q, \TV_{\rm bound})$.
\For{$y \in Y$} \Comment{Stage 1: Exponential-free centered-moment pass}
    \If{$J_y = \varnothing$}
        \State Set $(\abs{\tA_y}, \delta_y, \text{status}_y) \gets (0, 0, \texttt{exact\_zero})$
    \Else
        \State $(\abs{\tA_y}, \delta_y, \text{res}) \gets \texttt{CenteredMomentCert}(p, J_y, \alpha, z_y, r_{\max})$
        \State Set $\text{status}_y \gets \texttt{resolved}$ if $\text{res}$ else $\texttt{unresolved}$
    \EndIf
\EndFor
\State Update $I \gets \{y : \text{status}_y=\texttt{resolved}\}$, $U \gets \{y : \text{status}_y=\texttt{unresolved}\}$
\State $N \gets N_0$
\While{($I = \varnothing$ \textbf{or} $\max_{y\in I} h_y\abs{\tA_y}^\beta \le \max_{y\in U} h_y\delta_y^\beta$ \textbf{or} $\frac{B_U}{S_I+B_U} > \eta - \varepsilon_{\rm real}$) \textbf{and} $N \le N_{\max}$ \textbf{and} $U\ne\varnothing$}
    \For{$y \in U$} \Comment{Stage 2: Selective modular refinement on $U$ only}
        \If{$p^{-N} < \delta_y$}
            \State Evaluate $\tA_y \gets \sum_{j\in J_y} \alpha_j \expp(z_{yj}) \bmod p^N$ using cutoff \eqref{eq:cutoff}
            \State $\delta_y \gets p^{-N}$
            \If{$\abs{\tA_y} > \delta_y$}
                \State $\text{status}_y \gets \texttt{resolved}$; move $y$ from $U$ to $I$
            \EndIf
        \EndIf
    \EndFor
    \State Recompute $S_I \gets \sum_{y\in I} h_y\abs{\tA_y}^\beta$ and $B_U \gets \sum_{y\in U} h_y\delta_y^\beta$
    \State $N \gets N + \Delta N$
\EndWhile
\State Form exact conditional distribution $Q(y) = h_y\abs{\tA_y}^\beta / S_I$ on $I$ (and $0$ on $U$)
\State Compute normalized dyadic vector $\widehat Q$ at bit budget $b$ via Proposition~\ref{prop:dyadic}
\State \Return $\bigl(\argmax_{y\in I} h_y\abs{\tA_y}^\beta,\ \text{first maximizer},\ \widehat Q,\ \frac{B_U}{S_I+B_U} + \TV(Q,\widehat Q)\bigr)$
\end{algorithmic}
\end{algorithm}

\begin{proposition}[Eventual termination and correctness]\label{prop:eventual}
Assume $Y$ is finite, at least one amplitude $A_y\ne0$, every unresolved positive
radius tends to zero under refinement ($N_{\max}=\infty$), and $\varepsilon_{\rm real}<\eta$.
For $\beta\in\N$ and $h_y\in\Q_{>0}$, Algorithm~\ref{alg:hybrid} terminates in finitely many steps and certifies both the exact maximizer set and a total-variation bound at most $\eta$.
\end{proposition}

\begin{proof}
Every nonzero amplitude $A_y$ resolves as soon as $\delta_y<\abs{A_y}$. Since $Y$ is finite, after finitely many steps $I$ contains all nonzero coordinates and $U$ contains only coordinates with $A_y=0$. As $\delta_y\to 0$ on $U$, $B_U=\sum_{y\in U}h_y\delta_y^\beta\to 0$, so both \eqref{eq:rank} and $B_U/(S_I+B_U)\le\eta-\varepsilon_{\rm real}$ hold after finitely many steps. Proposition~\ref{prop:real} completes the proof.
\end{proof}

\subsection{Scale invariance and distinction from task accuracy}

\begin{proposition}[Scale invariance]\label{prop:scaling}
For any $c\in\Qp^\times$, scaling
$(A_y,\tA_y,\delta_y)\mapsto(cA_y,c\tA_y,\abs{c}\delta_y)$
preserves the resolved set $I$, the conditional distribution $Q$, the dominance condition \eqref{eq:rank}, and the TV bound \eqref{eq:weighted-tv}.
\end{proposition}

\begin{proof}
All norms and error radii scale by $\abs{c}$, so all real weights $w_y$ and upper bounds $h_y\delta_y^\beta$ scale by the common positive factor $\abs{c}^{\beta}$.
\end{proof}

Finally, we emphasize that \emph{numerical coverage} (\texttt{NumCov}, the fraction of instances whose exact top-ranked output is certified) and \emph{TV coverage} (\texttt{TVCov}, the fraction whose total-variation bound is at most $\eta$) measure agreement with the exact mathematical readout \eqref{eq:readout-intro}, not predictive accuracy against external ground-truth labels.

For instance, take $p=3$, $\beta=1$, $h\equiv 1$, and the six-instance family
$$
A^{(t)}=(3^t,3^{t+1},3^{t+2},0),
\qquad t=1,\ldots,6,
$$
on $Y=\{0,1,2,3\}$. For every $t$, the exact readout is $P_1=(9/13,3/13,1/13,0)$ with top-ranked output $0$. At common precision $3^{-N}$, instance $t$ resolves if and only if $t<N$, yielding $\mathrm{NumCov}_N=\min\{6,\max\{0,N-1\}\}/6$. If external task labels equal $1$ for $t\in\{2,5\}$ and $0$ otherwise, Table~\ref{tab:coverage} shows that empirical accuracy on the certified subset need not even be monotone in $N$.

\begin{table}[ht]
\centering
\caption{Exact separation between numerical certification coverage and external task accuracy for the synthetic family $A^{(t)}=(3^t,3^{t+1},3^{t+2},0)$, $t=1,\ldots,6$.}
\label{tab:coverage}
\small
\resizebox{\linewidth}{!}{%
\begin{tabular}{cccc}
\toprule
Precision $N$ & Numerical coverage (\texttt{NumCov}) & Overall model accuracy & Certified-subset accuracy\\
\midrule
1 & $0$   & $2/3$ & undefined\\
3 & $1/3$ & $2/3$ & $1/2$\\
5 & $2/3$ & $2/3$ & $3/4$\\
7 & $1$   & $2/3$ & $2/3$\\
\bottomrule
\end{tabular}%
}
\end{table}

%=====================================================================
\section{Reference implementation and numerical experiments}
\label{sec:implementation}
%=====================================================================

\subsection{Architecture and verification suite}

The reference implementation comprises two self-contained, standard-library Python modules: \texttt{padic\_certified.py} and \texttt{verify\_and\_benchmark.py}.
\begin{itemize}
\item \textbf{Exact rational and modular arithmetic:} For modular evaluation, the code factors powers of $p$ out of factorial denominators $n!$ and inverts only their unit parts modulo $p^N$, selecting truncation cutoffs via Lemma~\ref{lem:tail}. Coincident phases are merged algebraically prior to evaluation.
\item \textbf{Centered-moment engine:} Algorithm~\ref{alg:moment} is implemented in exact rational arithmetic (\texttt{fractions.Fraction}), selecting the optimal phase center via Proposition~\ref{prop:center-choice}. Because a moment representative approximates the unit-shifted amplitude $\expp(-b_y)A_y$, the software uses its certified norm information without intersecting its digit ball with unshifted modular representatives.
\item \textbf{Audit-ready certificate records:} Every readout evaluation outputs a structured record separating (i) per-output representatives, radii, and method tags (\texttt{modular}, \texttt{moment}, or \texttt{exact}), (ii) the sets $I$ and $U$, (iii) the certified maximizer set and top-ranked output, (iv) the exact rational law $Q$ and $b$-bit dyadic vector $\widehat Q$, and (v) the two-part TV error ledger.
\end{itemize}
The automated regression suite (\texttt{python verify\_and\_benchmark.py --tests --demo}) verifies Lemma~\ref{lem:tail}, heterogeneous-radius dominance (Example~\ref{ex:radii}), rounding-induced rank inversion (Example~\ref{ex:rounding}), three-source second-order resolution (Example~\ref{ex:three}), exact-zero cancellation in the benchmark workload (the output $y=3$ of Section~\ref{subsec:repro}), trajectory confinement, and zero-coupling degeneracy (Corollary~\ref{cor:degenerate}).

\subsection{Benchmark setup}

We evaluate our certification policies on $p=5$ over the $64$ leaf addresses of a depth-$3$ quaternary prefix tree:
\[
Y=
\left\{
a_0+5a_1+25a_2:
a_0,a_1,a_2\in\{0,1,2,3\}
\right\}.
\]
The source configuration and balanced coefficients are
\[
(x_1,x_2,x_3,x_4)=(0,1,5,6),
\qquad
(\alpha_1,\alpha_2,\alpha_3,\alpha_4)=(1,1,-1,-1),
\]
with free affine-chain phase $z(y,x)=5(y-x)^2$ ($k=1$, $\mu=10$, $\epsilon=1$, $a=1$, $V=0$).
We test two model configurations: an \emph{unmasked} workload ($J_y=\{1,2,3,4\}$ for all $y\in Y$) and a \emph{depth-one masked} workload retaining only sources with $\abs{y-x}\le 5^{-1}$.
The readout parameters are $\beta=1$, $h_y\equiv 1$, target TV tolerance $\eta=10^{-2}$, and $b=20$-bit dyadic rounding ($\varepsilon_{\rm real}=64/2^{21}\approx 3.05\times 10^{-5}\ll\eta$).

We compare four certification policies:
\begin{enumerate}
\item \emph{Fixed-low}: one-shot modular evaluation at $N=2$.
\item \emph{Fixed-high}: one-shot modular evaluation at $N=10$.
\item \emph{Adaptive}: modular evaluation starting at $N_0=2$ with step $\Delta N=2$ and cap $N_{\max}=24$, refining only unresolved outputs $y\in U$.
\item \emph{Hybrid} (Algorithm~\ref{alg:hybrid}): centered-moment tests up to order $r_{\max}=3$, followed if needed by adaptive modular evaluation ($N_0=2$, $\Delta N=2$, $N_{\max}=24$).
\end{enumerate}

We report \texttt{Mean} $N=\frac{1}{|Y_+|}\sum_{y:\delta_y>0}(-\log_p\delta_y)$ (the average effective precision exponent across positive-radius representatives), \texttt{Terms} (cumulative modular Taylor terms evaluated), \texttt{MomTerms} ($\sum r|J_y|$ across tested moment orders), and wall-clock time across $5$ repetitions (\texttt{python verify\_and\_benchmark.py --benchmark --repeats 5}). Note that \texttt{Mean} $N$ mixes two radius types---modular truncation radii $p^{-N}$ and centered-moment remainder radii $\varepsilon_r=p^{-\rho_r(s_D)}$---so it is a per-output progress diagnostic rather than a uniform notion of precision; equal exponents of the two types do not imply equal certified quality.

\subsection{Benchmark results and overhead analysis}

Tables~\ref{tab:benchmark-unmasked} and~\ref{tab:benchmark-masked} summarize the results for the unmasked and masked four-source workloads.

\begin{table}[htbp]
\centering
\caption{Unmasked four-source workload ($|Y|=64$, $p=5$, $\eta=10^{-2}$, $20$-bit dyadic output).
Times are means $\pm$ sample standard deviations over five repetitions.}
\label{tab:benchmark-unmasked}
\small
\setlength{\tabcolsep}{5pt}
\renewcommand{\arraystretch}{1.15}
\resizebox{\linewidth}{!}{%
\begin{tabular}{@{}lrrrrrc@{}}
\toprule
Policy & NumCov & TVCov & Mean $N$ & Terms & MomTerms & Time (s) \\
\midrule
Fixed-low  & $0.0\%$   & $0.0\%$   & $2.00$  & $440$  & $0$
& $0.002429 \pm 0.000041$ \\
Fixed-high & $100.0\%$ & $100.0\%$ & $10.00$ & $2476$ & $0$
& $0.003983 \pm 0.000217$ \\
Adaptive   & $100.0\%$ & $100.0\%$ & $4.00$  & $1304$ & $0$
& $0.005448 \pm 0.000234$ \\
Hybrid     & $100.0\%$ & $100.0\%$ & $3.33$  & $0$    & $5472$
& $0.026059 \pm 0.000991$ \\
\bottomrule
\end{tabular}%
}
\end{table}

\begin{table}[htbp]
\centering
\caption{Depth-one masked four-source workload ($\abs{y-x}\le 5^{-1}$, $|Y|=64$, $p=5$, $\eta=10^{-2}$, $20$-bit dyadic output).}
\label{tab:benchmark-masked}
\small
\setlength{\tabcolsep}{5pt}
\renewcommand{\arraystretch}{1.15}
\resizebox{\linewidth}{!}{%
\begin{tabular}{@{}lrrrrrc@{}}
\toprule
Policy & NumCov & TVCov & Mean $N$ & Terms & MomTerms & Time (s) \\
\midrule
Fixed-low  & $0.0\%$   & $0.0\%$   & $2.00$  & $64$  & $0$
& $0.000860 \pm 0.000012$ \\
Fixed-high & $100.0\%$ & $100.0\%$ & $10.00$ & $220$ & $0$
& $0.001498 \pm 0.000111$ \\
Adaptive   & $100.0\%$ & $100.0\%$ & $4.50$  & $208$ & $0$
& $0.002844 \pm 0.000117$ \\
Hybrid     & $100.0\%$ & $100.0\%$ & $6.62$  & $0$   & $384$
& $0.003593 \pm 0.000149$ \\
\bottomrule
\end{tabular}%
}
\end{table}

Three key insights emerge from Tables~\ref{tab:benchmark-unmasked}--\ref{tab:benchmark-masked}: \emph{(i) Precision threshold vs.\ over-provisioning.} Fixed-low ($N=2$) fails because cancellation pushes $v_*$ to $2$ or higher, requiring $N\ge v_*+1\ge 3$ (Proposition~\ref{prop:sharp}); Fixed-high ($N=10$) succeeds but over-provisions by more than a factor of two relative to Adaptive ($\text{Mean }N=4.00$ unmasked, $4.50$ masked), on a per-output average of effective precision. \emph{(ii) Precision savings vs.\ Python loop overhead.} In the unmasked workload, Adaptive nearly halves the exponential Taylor terms ($1304$ vs.\ $2476$), yet its multi-pass bookkeeping makes it slightly slower in wall-clock time ($5.45$\,ms vs.\ $3.98$\,ms) than a single pass at $N=10$; when kernel evaluations dominate (large $k$ or high-degree potentials), term-count reduction governs runtime. \emph{(iii) Exponential-free certification.} Hybrid achieves $100\%$ certification with $\texttt{Terms}=0$, resolving every output via moments of order $r\le 3$; consistent with Section~\ref{subsec:crossover}, exact rational arithmetic (\texttt{fractions.Fraction}) incurs higher Python overhead ($26.1$\,ms unmasked, $3.59$\,ms masked) than native integer arithmetic modulo $5^N$.

\subsection{Coverage, ties, and reproducibility}\label{subsec:repro}

Three further points sharpen the interpretation and bear on reproducibility. \emph{Coverage is measured per input:} since each workload uses one input, $\texttt{NumCov}$ and $\texttt{TVCov}$ are $0\%$ or $100\%$; sweeping the coefficient scale $5^t$ ($t=0,\ldots,5$) against fixed precision $N$ recovers the sharp threshold, with the first certified instance at $N=v_*+1$ ($N=3$ unmasked, $N=4$ masked; Table~\ref{tab:coverage-vs-N}). \emph{Exact ties are certified:} the unmasked workload has a $48$-way tie at norm $5^{-2}$ and one exactly zero output $y=3$ (recorded with radius $0$); the masked workload has a $24$-way tie at $5^{-3}$. The certificate returns the full maximizer set, and the regression suite checks agreement with the exact tie set. \emph{Reproducibility:} timings are single-threaded wall-clock means $\pm$ sample standard deviations, while the deterministic metrics are machine-independent; the implementation records the Python version and platform in the benchmark JSON output.

\begin{table}[htbp]
\centering
\caption{Coverage versus absolute precision on the scaled four-source workloads. Each entry is the number of the six coefficient scales $5^t$ ($t=0,\ldots,5$) whose exact top-ranked output is certified at fixed precision $N$; the staircase rises at the sharp threshold $N=v_*+1$.}
\label{tab:coverage-vs-N}
\small
\resizebox{\linewidth}{!}{%
\begin{tabular}{@{}lrrrrrrr@{}}
\toprule
Precision $N$ & $2$ & $3$ & $4$ & $5$ & $6$ & $7$ & $8$ \\
\midrule
Unmasked ($v_*=2$) & $0/6$ & $1/6$ & $2/6$ & $3/6$ & $4/6$ & $5/6$ & $6/6$ \\
Masked ($v_*=3$)   & $0/6$ & $0/6$ & $1/6$ & $2/6$ & $3/6$ & $4/6$ & $5/6$ \\
\bottomrule
\end{tabular}%
}
\end{table}

\subsection{Two-source closed-form speedup}

For the balanced two-source workload $(x_1,x_2)=(0,1)$, $(\alpha_1,\alpha_2)=(1,-1)$, the isometric identity of Theorem~\ref{thm:two} turns the aggregate into the closed form of Corollary~\ref{cor:two-free}, eliminating the exponential series entirely. The resulting speedup---$4.8\times$ at the minimal sufficient precision $N=5$, rising to $87.2\times$ at $N=50$---is tabulated in the supplementary material.

%=====================================================================
\section{Concluding remarks}\label{sec:conclusion}
%=====================================================================

We have developed a rigorous finite-precision certification framework for norm-based readouts of $p$-adic amplitudes and discriminants. Because the $p$-adic absolute value is locally constant away from zero, discrete decisions---including exact ties in both amplitude maximization and discriminant minimization---can be certified with mathematical exactness at sharp integer precision thresholds ($v_*+1$ and $v^{(2)}+1$, respectively). For exponential kernel aggregates, our centered-moment certificate algebraically captures leading-order phase cancellation prior to series truncation, enabling exponential-free norm certification with optimal finite center selection. Applied to discrete path-integral actions along uniform affine chains, these tools reveal explicit action centers and exact output-independence at zero effective quadratic coupling.

Promising directions for future research include: (i) extending centered-moment remainder bounds to multivariate and higher-degree polynomial potentials in discrete $p$-adic path integrals \cite{HuKim2026}; (ii) accelerating moment evaluation via fast multipole or hierarchical prefix-tree recurrences so that $m_n(b_y)$ is shared across subtrees rather than recomputed per output; and (iii) coupling the certified discriminant inference layer of Section~\ref{sec:discriminants} with Hensel-lifting and combinatorial optimization algorithms for certified $p$-adic model training \cite{Martins}.

\section*{Acknowledgements}
This manuscript originated from the authors' long-term conversations and discussions with AI assistants, which also provided useful suggestions and assistance in improving the manuscript. The authors are fully responsible for the content of the manuscript.

 \section*{Data availability}
All data supporting the findings of this study are available within the paper and its Supplementary Information. The benchmark results are reported in the tables of the paper, and the underlying certificate records, coverage counts, term counts, and timings are produced by the reference implementation (\texttt{padic\_certified.py}), the verification/benchmark suite (\texttt{verify\_and\_benchmark.py}), and the two-source table generator (\texttt{table5\_two\_source.py}), all of which are provided as Supplementary Information and reproduce every table and certificate record reported in the paper. No external datasets were used; all data are synthetic and are generated deterministically by the provided code.

\section*{Code availability}
The code supporting this study is available in the Supplementary Information as three self-contained, standard-library Python~3 modules: \texttt{padic\_certified.py} (the certification library), \texttt{verify\_and\_benchmark.py} (the regression and benchmark suite, run via \texttt{python verify\_and\_benchmark.py --tests --demo} and \texttt{python verify\_and\_benchmark.py --benchmark --repeats 5}), \\and \texttt{table5\_two\_source.py} (the two-source table generator).

\end{document}